\documentclass[fleqn,usenatbib]{mnras}

\usepackage{newtxtext,newtxmath}
\usepackage[T1]{fontenc}
\usepackage{ae,aecompl}
\usepackage{graphicx}
\usepackage{amsmath}
\usepackage{amssymb}
\usepackage[mathscr]{euscript}

\let\mathscr\relax 
\usepackage[scr]{rsfso}
\usepackage{enumitem}
\usepackage{tabularx, booktabs, makecell}
\usepackage{siunitx}
\newcommand{\daa}{\Delta\alpha/\alpha}

\DeclareMathAlphabet{\mathpzc}{OT1}{pzc}{m}{it}

\title[Information Criterion for Spectroscopy]{Getting the model right; an information criterion for spectroscopy}

\author[J. K. Webb et al]{
John K. Webb$^1$\thanks{E-mail: jkw@phys.unsw.edu.au},
Chung-Chi Lee$^{2}$\thanks{E-mail: lee.chungchi16@gmail.com},
Robert F. Carswell,$^{3}$
and Dinko Milakovi{\'c}$^{4}$. \\\\
$^{1}$School of Physics, University of New South Wales Sydney, NSW 2052, Australia\\
$^{2}$DAMTP, Centre for Mathematical Sciences, University of Cambridge, Cambridge CB3 0WA, UK\\
$^{3}${Institute of Astronomy, Madingley Road, Cambridge CB3 0HA, U.K.}\\
$^{4}${European Southern Observatory, Karl-Schwarzschild-str 2, 85748 Garching bei M{\"u}nchen, Germany}
}

\date{Accepted ... Received ...; in original form 15 September, 2020}
\pubyear{2020}
\begin{document}
\label{firstpage}
\pagerange{\pageref{firstpage}--\pageref{lastpage}}
\maketitle

\begin{abstract}
Robust model-fitting to spectroscopic transitions is a requirement across many fields of science. The corrected Akaike and Bayesian information criteria (AICc and BIC) are most frequently used to select the optimal number of fitting parameters. In general, AICc modelling is thought to overfit (too many model parameters) and BIC underfits. For spectroscopic modelling, both AICc and BIC lack in two important respects: (a) no penalty distinction is made according to line strength such that parameters of weak lines close to the detection threshold are treated with equal importance as strong lines and (b) no account is taken of the way in which spectral lines impact on narrow data regions. In this paper we introduce a new information criterion that addresses these shortcomings, the {\it Spectral Information Criterion} (SpIC). Spectral simulations are used to compare performances. The main findings are (i) SpIC clearly outperforms AICc for high signal to noise data, (ii) SpIC and AICc work equally well for lower signal to noise data, although SpIC achieves this with fewer parameters, and (iii) BIC does not perform well (for this application) and should be avoided. The new method should be of broader applicability (beyond spectroscopy), wherever different model parameters influence separated small ranges within a larger dataset and/or have widely varying sensitivities.
\end{abstract}


\section{Introduction}

In recent papers, we described new automated spectral modelling algorithms: GVPFIT, \cite{Bainbridge2017,gvpfit17} and AI-VPFIT, \citet{Lee2020AI-VPFIT}. These use artificial intelligence techniques to create multi-parameter models of complex absorption systems. In evolutionary algorithms of this sort, the fittest model from the current generation is selected from a set of candidates, becoming the parent for the succeeding generation. As model complexity develops, an information criterion (IC) is used in each generation to select the fittest model. The choice of information criterion defines the evolutionary path through parameter space and influences the outcome of the final model. 

In the works cited above and in \citet{MilakovicHE05152020}, we focused on the widely used corrected Akaike Information Criterion, AICc, \citep{Akaike1974,Hurvich1989},
\begin{equation}\label{eq:AICc}
\mathrm{AICc} = \chi^2 + \frac{2 k N }{N - k -1}\,\,,
\end{equation}
but also experimented with the Bayesian Information Criterion, BIC \citep{Bozdogan1987},
\begin{equation}\label{eq:BIC}
\mathrm{BIC} = \chi^2 + k\ln(N).
\end{equation}
Additional derivations can be found in \citet{Portet2020} (AICc) and \cite{Bhat2010} (BIC). In both AICc and BIC, $k$ is the total number of free model parameters and $N$ is the number of data points. AICc and BIC are used as comparison baselines since they are so widely applied. We do not consider AIC itself since AICc is more appropriate for $N/k\!<\!40$ \citep{Burnham2002} and asymptotes to AIC for $N\!\!\gg\!\!k$. The second term in each criterion is a penalty function that increases with increasing $k$, so the information criterion minimises at the best fit, avoiding too many free parameters in the final model. Although both AICc and BIC comply with the ``principle of parsimony'', they do so in different degrees. 

The principle of parsimony\footnote{Also referred to as ``Occam's razor'' (14th century) although the concept dates back to at least Aristotle. Newton (orig. in Latin): ``We are to admit no more causes of natural things than such as are both true and sufficient to explain their appearances'' \citep{Newton1726}. Einstein: ``... the grand aim of all science ... is to cover the greatest possible number of empirical facts by logical deduction from the smallest possible number of hypotheses or axioms'', \citep{Einstein1954}. \cite{Laird1919} provides a comprehensive historical account.} aims to strike a balance between parameter variance and model bias. As the number of free parameters in the model increases, so do the estimated parameter errors. On the other hand, as one approaches the true number of model parameters, bias decreases. Since different information criteria impose differing penalties, they also permit different levels of residual bias in the final model. The AICc penalty is smaller than that of BIC, so best-fit models based on AICc naturally comprise more model parameters relative to BIC. A basic requirement in any model fitting is to minimise that bias. \citet{Burnham2002} provide comprehensive discussions on these points and on many aspects of model fitting and \citet{Liddle2007} describes the application of several information criteria to various problems in astrophysics.

In spectroscopic applications, some model parameters may impact only over a very small subset of the overall dataset being fitted. In this paper we consider the example of a well-known quasar absorption system comprising multiple narrow components, many closely blended, spread over a large observed wavelength range. In constructing a model, one needs to assess how many model components are needed. When using an information criterion to decide whether any particular trial component is needed, specifying the number of data points associated with the trial component is not straightforward. How far out from the line centre does the profile extend? Is it reasonable to assume that each individual spectral line spans the entire data region being modelled? 

A further problem arises for line strength. The AICc penalises each model parameter equally, as does the BIC. In a spectroscopic context, this means that a very weak spectral line, potentially close to the detection threshold, has an identical penalty term to a strong feature. The consequence of that (in absorption line spectroscopy where there are multiple spectral features) is that the model can end up with too many parameters, particularly multiple weak lines.

These two apparent inadequacies inherent to AICc and BIC (in the context considered here at least) suggest that a new type of penalty function could be useful. 
In Section \ref{SpIC}, we introduce a new ``hybrid'' information criterion, SpIC, that allows for strongly localised parameter impact and absorption line strength variation. It also strikes a balance between the relative overfitting of AICc and underfitting of BIC. Section \ref{Synths} describes a set of synthetic spectra used to assess the each information criterion performs (discussed in Section \ref{Analysis}) and finally Section \ref{Results} presents the full set of numerical results and list the main conclusions reached.

\section{A penalty function based on spectral line strengths}\label{SpIC}

\begin{figure}
\centering
\includegraphics[width=0.8\linewidth]{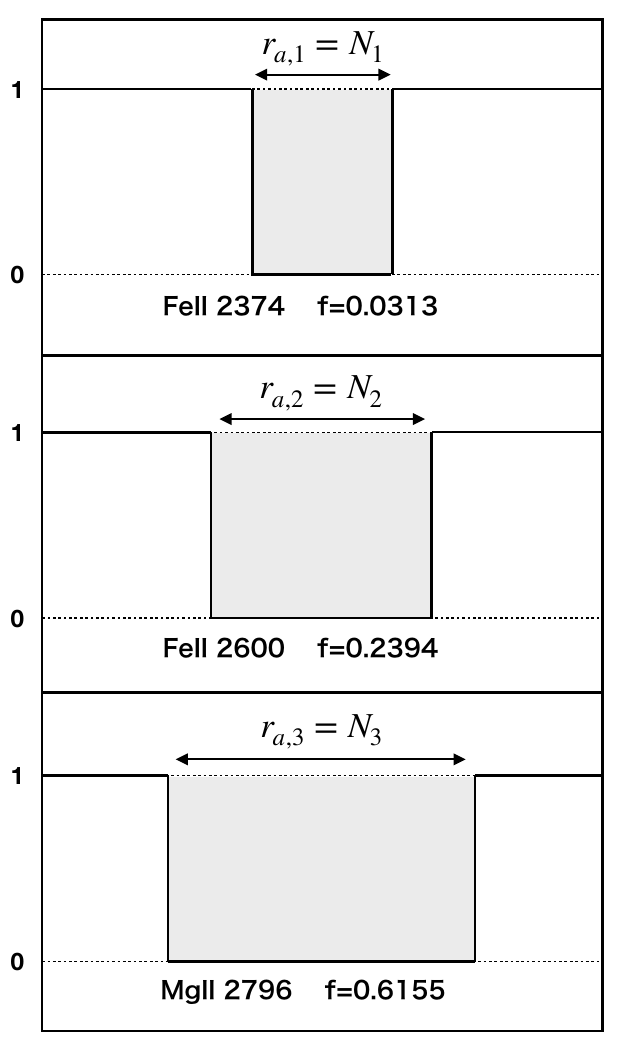}
\caption{Interpretation of $r_{a,j}$ and hence $R_a$. Idealised absorption profiles are represented here as rectangles. The widths depend on the column density, oscillation strength and line broadening parameter, $b$. We illustrate 3 cases, FeII 2374, FeII 2600 and MgII 2796. The number of pixels that each absorption line can influence (the grey shaded region), can be interpreted as $r_{a,j}$, defined in Eq.~\eqref{eq:Ra}, if the error array $\sigma$ equals 1 everywhere.
\label{fig:rec_example}}
\end{figure}

We introduce a new information criterion, SpIC, with the following aims. First, we want to investigate whether it is advisable to strike a balance between the {\it relative} overfitting tendency of AICc and the underfitting of BIC. Second, the new penalty function still depends on only 2 quantities, but instead of $k$ and $N$, we make use of the number of model parameters per spectroscopic line and the strength of each spectroscopic line. Both AICc and BIC treat all model parameters as being of equal importance. This is inappropriate for spectroscopic applications in which there is a wide range in absorption line strengths, with some components close to the detection threshold. Third, an individual spectral line may be narrow compared to the spectral range being fitted.

\subsection{Definition of SpIC}

We now define the new spectroscopic information criterion, which is
\begin{equation}
\label{eq:SpIC}
\mathrm{SpIC} = \chi^2 + \sum_{a=1}^Q \left[ \frac{2 f k_a R_a }{R_a - k_a -1} + (1-f) k_a\ln(R_a) \right]\,\,,
\end{equation}
where $Q$ is the total number of velocity components in the model, and
\begin{equation} \label{eq:chisq}
\chi^2 = \sum_{j=1}^M \sum_{i=1}^{N_j} \frac{(\mathcal{I}_{i,j}^\mathrm{data} - \mathcal{I}_{i,j}^\mathrm{model})^2}{ \sigma^2_{i,j}}\,\,,
\end{equation}
where $\mathcal{I}_{i,j}^\mathrm{data}$, $\mathcal{I}_{i,j}^\mathrm{model}$, and $\sigma_{i,j}$ are the normalised observed spectral intensity, the model fit, and the error array on $\mathcal{I}_{i,j}^\mathrm{data}$ for the $\it i^{th}$ pixel in $j^{th}$ spectral segment, and $M$ is the number of segments comprising the whole dataset being modelled. Each spectral segment has $N_j$ pixels such that $N\!=\!\sum^M \!N_j$.

$R_a$ has replaced the usual dependence on $N$ in the standard AICc and BIC and instead allows for the relative strength of each spectral line. $k_a$ is the number of free parameters associated with each component in the absorption complex such that $k\!=\!\sum^Q \!k_a$. The value of $f$ ($0\!\leqslant\!f\!\leqslant\!1$) can be used to fine-tune performance, although here the only non-extreme value used is 1/2, which is found to work well, as discussed shortly.

The characteristics we require of $R_a$ are: (i) $R_a$ increases with increasing absorption line strength, (ii) it should properly account for the spectral error array, such that spectral regions with poor signal to noise are weighted accordingly, and (iii) a general requirement is to fit multiple species simultaneously. These are provided to the modelling algorithm as $M$ separate spectral segments. Therefore, we define
\begin{equation} \label{eq:Ra}
R_a = \sum_{j=1}^M r_{a,j},
\end{equation}
where the quantities $r_{a,j}$ parameterise both the effective number of pixels and line strength information for each absorption profile in each segment of the model.

\subsection{Interpreting $R_a$}
\label{subsec:Ra}

Consider the following idealised example. Suppose each absorption line is a simple rectangle, with zero intensity at its base (Fig. \ref{fig:rec_example}). Suppose further that $\sigma_{i,j} = 1$ everywhere. Then $r_{a,j}$ equates to the effective number of pixels that one absorption line can influence. In this case, $R_a$ becomes the effective number of pixels corresponding to one velocity component, for one atomic species. Finally, when the summation over $Q$ is carried out in Eq. \eqref{eq:SpIC}, $R_a$ leads to the effective number of pixels influenced by all velocity components and all atomic species. In this simple example we can thus see the close analogy between $R_a$ and $N$.

\subsection{Defining $r_{a,j}$}

The above case is idealised and a real absorption line shape is obtained from a Voigt profile convolved with the appropriate instrumental profile. Pixels closer to the line centre provide more information, so need higher weighting. 
Moreover, spectral signal to noise varies from pixel to pixel and this also requires appropriate weighting. Therefore, for each velocity component (and for each spectral segment), we define $r_{a,j}$ as 
\begin{equation} \label{eq:smallra}
r_{a,j} = \sum_{i=1}^{N_j} \left( \frac{ 1-I_{i,j}^a}{ \sigma_{i,j}}\right),
\end{equation}
i.e. we calculate the normalised depth at each pixel, summed over all pixels contributing to that velocity component (i.e. redshift). We have assumed normalised data i.e. a continuum of unity. The quantity $a$ indicates the $a^{th}$ velocity position in the absorption system (at which several transitions reside). The individual profile intensities and the overall model intensity are related by
\begin{equation}
\mathcal{I}_{i,j}^\mathrm{model} = \prod_{a=1}^{Q} I_{i,j}^a \,\,.
\end{equation}

\subsection{Penalty function for SpIC}

\begin{figure}
\centering
{\includegraphics[width=0.98\linewidth]{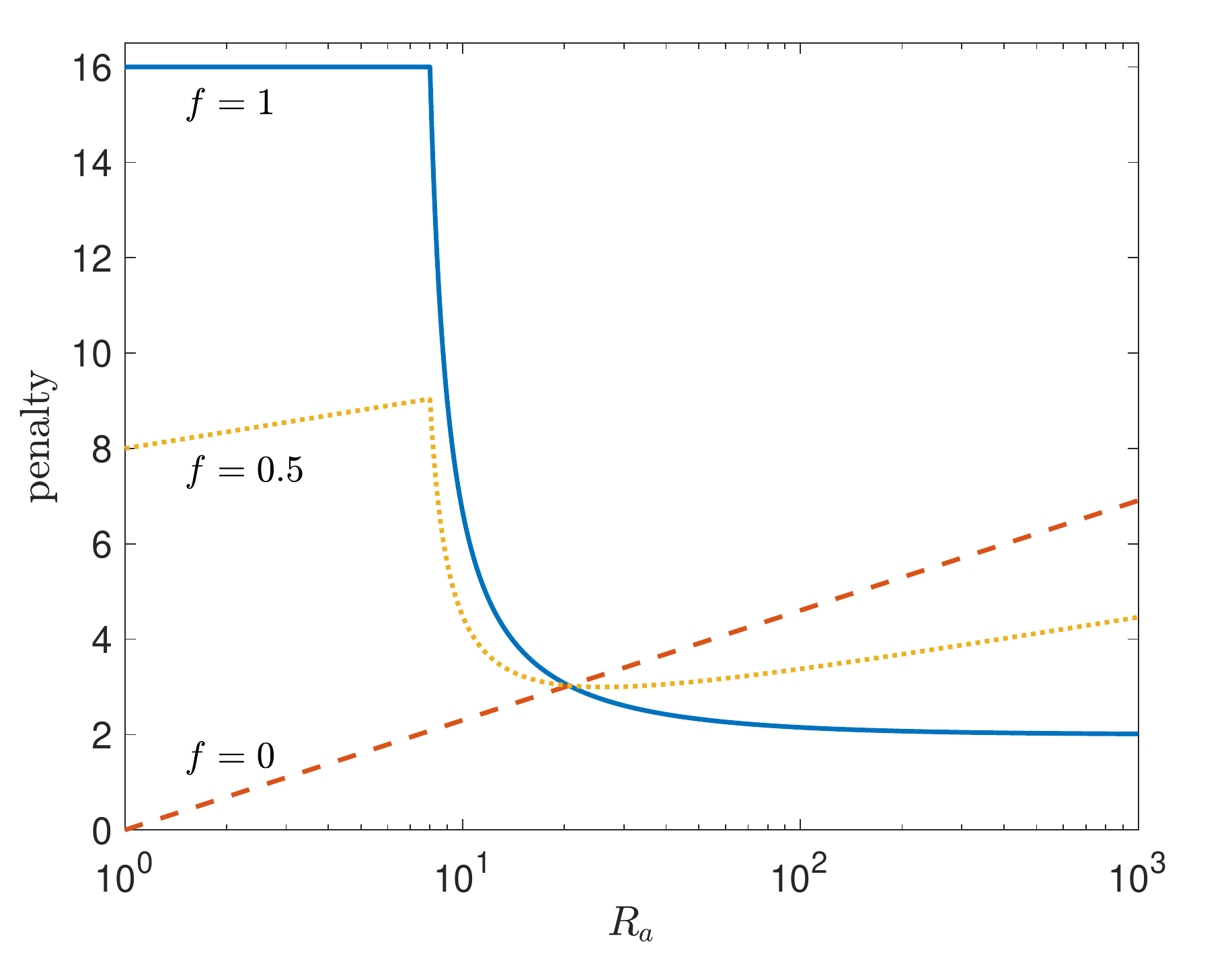}}
{\caption{These curves illustrate the quantity inside the square brackets of Eq.\eqref{eq:SpIC}, for $k_a=6$ (the value of 6 is explained in Section \ref{SpIC}). The solid blue, dotted yellow, and dashed red curves correspond to $f=1$, 0.5 and 0, respectively.
\label{fig:pen-Ra}}}
\end{figure}

Fig. \ref{fig:pen-Ra} illustrates the quantity inside the square brackets of Eq.\eqref{eq:SpIC} with $k_a\!=\!6$. The reason for $k_a\!=\!6$ is specific to the application described in Section \ref{Synths}; synthetic spectra are used, based on 3 atomic species (MgI, MgII, FeII) using a total of 6 free parameters per absorption component (3 column densities, one velocity dispersion parameter, one temperature parameter, and one redshift). The figure illustrates how the penalty function behaves as a function of line strength, $R_a$. The first term within the brackets blows up if we allow $R_a \!=\! k_a+1$. Therefore a lower bound is imposed, $R_a\!\geq\!k_a+2$, forming the upper plateau in the solid blue curve (which illustrates $f\!=\!1$). The solid blue curve may be considered the AICc analogue. The dashed red line corresponds to $f\!\!=\!\!0$, i.e. it is the BIC analogue. The blue curve thus shows how the penalty is large for weak absorption lines and small for strong lines, suggesting a tendency to underfit weak absorption lines and overfit strong lines. Conversely, the red curve illustrates a very small penalty for weak absorption lines and a high penalty for strong absorption lines, suggesting a tendency to overfit weak lines and underfit strong lines. Alone, neither satisfy an appropriate balance. The hybrid illustrated by the dotted yellow line, using $f\!\!=\!\!1/2$, forms a compromise penalty.

We consider 3 cases of Eq.\eqref{eq:SpIC}:

\noindent{\bf Case I, ($f\!=\!1$):} analogous to AICc, where the usual dependence on $N$ has been replaced with a dependence on the line strength measure given by Eq.\eqref{eq:Ra} and the number of free model parameters per spectra region is $k_a$. We denote this SpIC$_{\textrm{A}}$.

\noindent{\bf Case II, ($f\!=\!1/2$):} equal weight is given to both terms in the square brackets of Eq.\eqref{eq:SpIC} i.e. we create a hybrid of the AICc and BIC analogues. We denote this SpIC$_{\textrm{H}}$.

\noindent{\bf Case III, ($f\!=\!0$):} analogous to BIC, where $N$ and $k$ have been replaced by Eq.\eqref{eq:Ra} and $k_a$. We denote this SpIC$_{\textrm{B}}$.

\section{Synthetic spectra} \label{Synths}

The $z_{{\textrm{abs}}}=1.15$ absorption complex in the spectrum of the bright quasar HE0515-0414 has been well studied. It is a complicated system spanning an unusually large redshift range, $1.14688 < z_{{\textrm{abs}}} < 1.15176$, corresponding to a velocity range $\sim\!\!700$ km/s. It comprises a range of line strengths and transitions from many species are detected. Here, for simplicity, we make use of only four species, MgI, MgII, FeII, and MnII and create simulated spectra based on an AI-VPFIT model of the $z_{{\textrm{abs}}}=1.15$ complex \citep{MilakovicHE05152020,Lee2020AI-VPFIT}. 

\begin{figure*}
\centering
\includegraphics[width=1.0\linewidth]{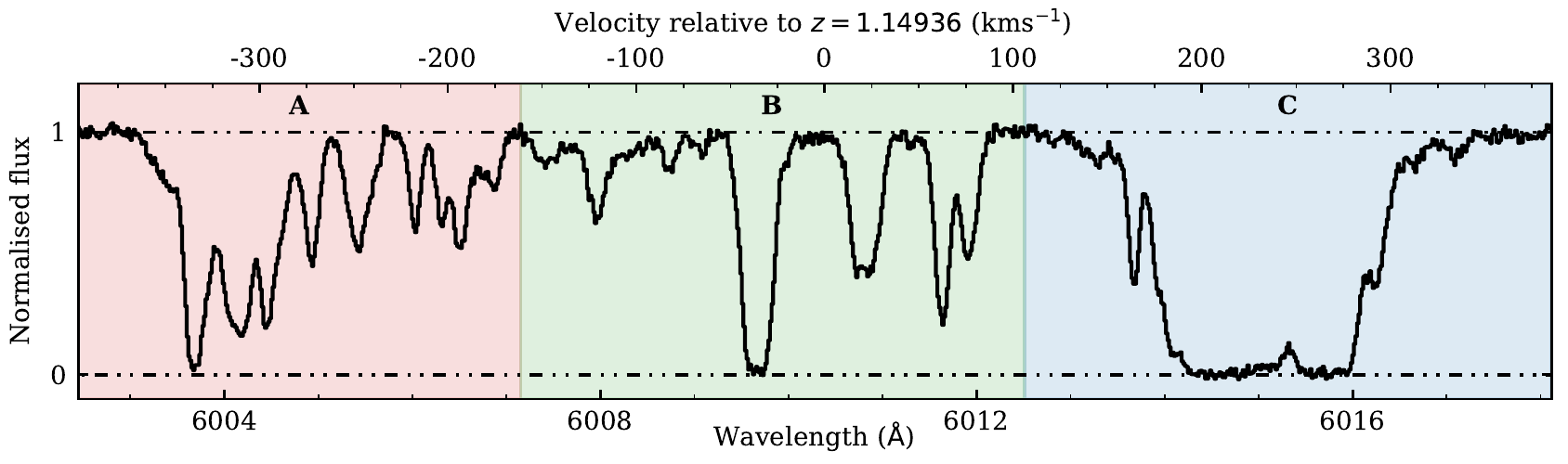}
\caption{Observed MgII 2796{\AA} profile for the $z_{abs}=1.15$ complex towards HE0515-0414. The spectral simulations are based on this absorption system. As described in Section \ref{Synths}, we split the data into 3 segments for modelling. Eight transitions are modelled simultaneously -- we only illustrate one transition here. The observations were obtained using HARPS on the ESO 3.6m telescope and are wavelength calibrated using a laser frequency comb. Full details are given in \citet{MilakovicHE05152020}.
\label{fig:realMg2796} }
\end{figure*}

The whole 700 km/s complex is divided into separate regions, along the lines of \cite{Kotus2017} and \cite{MilakovicHE05152020}. In the latter, 5 separate regions were defined and modelled independently (for reasons explained in that paper). Here it is more suitable to combine their regions I+II and III+IV, leaving 3 spectral segments, as did \cite{Kotus2017}, which we call A, B, and C. The real observational data is illustrated in Fig. \ref{fig:realMg2796}.

We simulate 8 transitions: MgI2853, MgII2796, 2803, FeII2344, 2374, 2383, 2586, 2600{\AA} for Regions A and B. Region C comprises higher column density transitions. For this region only we simulate the same 8 transitions but also add MnII2577, 2595, and 2606{\AA}. In generating the synthetic spectra, we set the parameter $\daa = \left(\alpha_z - \alpha_0\right)/\alpha_0 = 5 \times 10^{-6}$, where $\alpha_z$ is the average value of the fine structure constant in the $z_{{\textrm{abs}}}=1.15$ complex and $\alpha_0$ is the terrestrial value. 

We use Voigt profiles convolved with a Gaussian instrumental profile with $\sigma_{{\textrm{res}}} = 1.11$ km/s and a pixel size of $0.83$ km/s. The resolution dispersion, and signal to noise ratio per pixel (S/N)  values used correspond to those of the HARPS instrument on the ESO 3.6m telescope and are also representative of forthcoming data from existing facilities such as ESPRESSO on the VLT and future ones such as HIRES on the ELT. Two sets of spectra are generated, at S/N of 50 and 100. The noise is taken to be Gaussian. We thus create a total of 6 sets of spectra (three spectral regions and two S/N).

\begin{figure}
\centering
\includegraphics[width=1.02\linewidth]{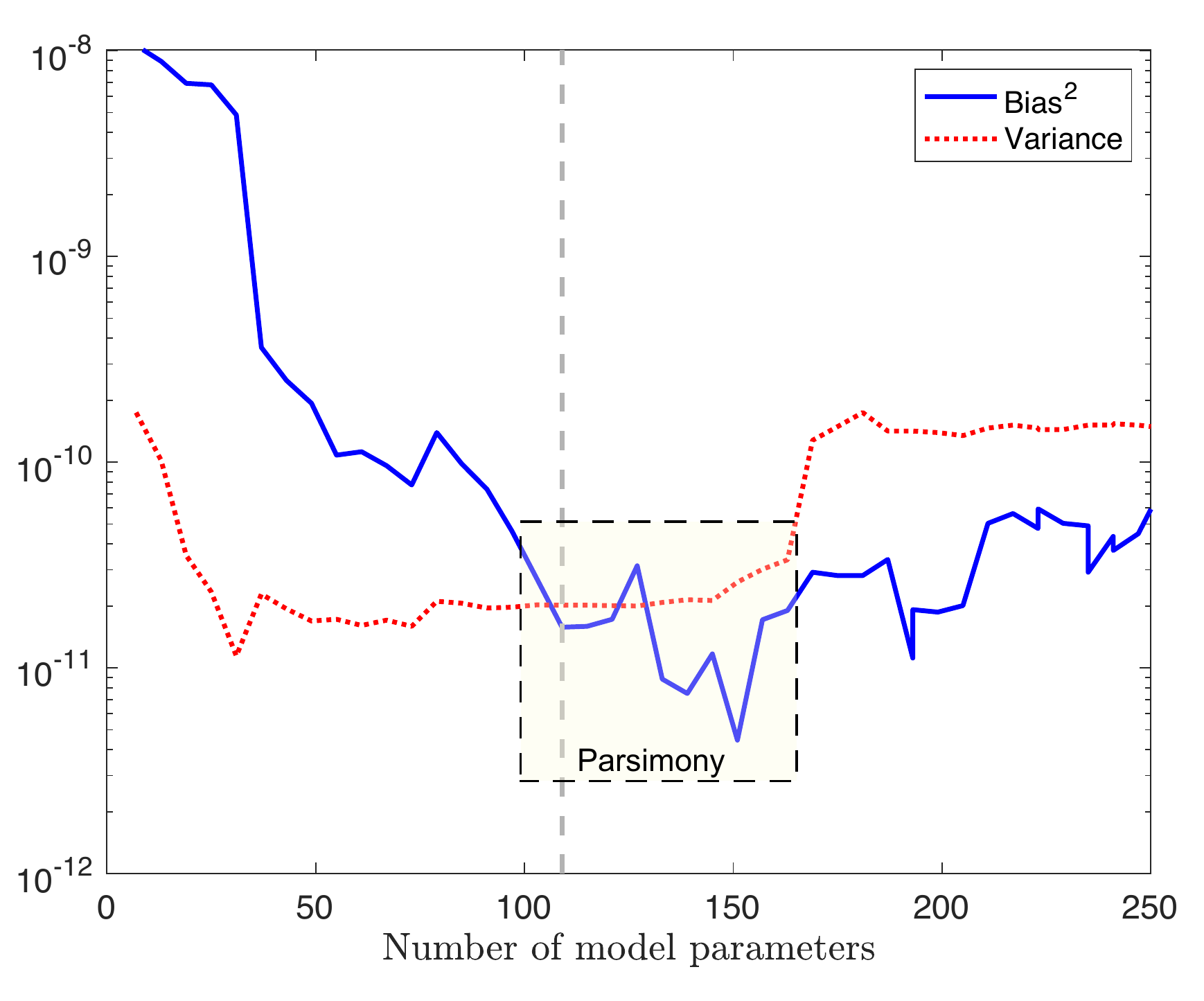}
\caption{Illustration of the parameter bias and variance relations for the fine structure constant, as a function of number of model parameters. The fiducial spectrum from Region B was used, with a signal to noise per spectral pixel of 50. The fiducial was derived using SpIC$_{\textrm{H}}$. Blue continuous line: bias$^2 = (\daa \rvert_{{\textrm{obs}}} - \daa\rvert_{{\textrm{true}}})^2$. Red continuous line: variance = $\sigma(\daa\rvert_{{\textrm{obs}}})^2$. Vertical dashed line: true number of parameters. The yellow shaded box illustrates an approximate region of parsimony. \label{fig:bias-variance} }
\end{figure}

\section{Analysis} \label{Analysis}

\subsection{Bias and variance} \label{Bias}

The trade-off between bias and variance, i.e. parsimony, can be seen by compiling results from each generation of an AI-VPFIT analysis \citep{Lee2020AI-VPFIT}. In Fig.\ref{fig:bias-variance}, the blue continuous line shows the bias$^2 = (\daa \rvert_{{\textrm{obs}}} - \daa\rvert_{{\textrm{true}}})^2$. When very few parameters are used to fit such a complex absorption system, the measured $\alpha$ is far from the true value, $\daa = \left(\alpha_z - \alpha_0\right)/\alpha_0 = 5 \times 10^{-6}$. As model complexity approaches the true number of parameters ($109$), bias decreases. Interestingly, within the empirical scatter, bias is approximately flat over a surprisingly large range in parameters ($\sim100-200$), broadly consistent with the effect seen in Fig. 7 of \citet{Bainbridge2017}. As the data become increasingly overfitted, bias on $\daa$ increases. This is expected because forcing additional velocity components into the model displaces lines from their correct positions, necessarily pushing $\daa$ away from the correct solution.

The red dotted line in Fig.\ref{fig:bias-variance} illustrates the variance, as obtained from the best fit Hessian matrix at each generation. In the overfitting region (i.e. $>109$ parameters), the parameter of interest is more poorly determined. Variance decreases with decreasing parameters, until the model becomes sufficiently simple as to be effectively meaningless (thus variance begins to increase again). Again, as with bias, there is a relatively broad region over which variance is approximately flat. These relatively flat regions in bias and variance are useful for $\daa$ measurement because there is a relatively low sensitivity to deviations from the true model, i.e. model non-uniqueness appears not to generate substantial additional uncertainty.

As discussed above, Fig. \ref{fig:bias-variance} shows that bias does not simply continue to decrease as more parameters are added, but in fact goes through a minimum before rising again. Physically this makes sense as additional spurious parameters would ultimately offset absorption components from their correct positions. The variance exhibits a similar behaviour (which is also expected for similar reasons). These properties raise an interesting question concerning how one invokes the principle of parsimony. In the application in this paper, since the minima are broad, the effect is unimportant. However, if the minima had been somewhat narrower such that the bias and variance minima were well separated, it would not be obvious how to apply parsimony and simultaneously minimise both.

\subsection{Comparing 5 information criteria} \label{5criteria}

\begin{figure*}
\centering
\includegraphics[width=0.98\linewidth]{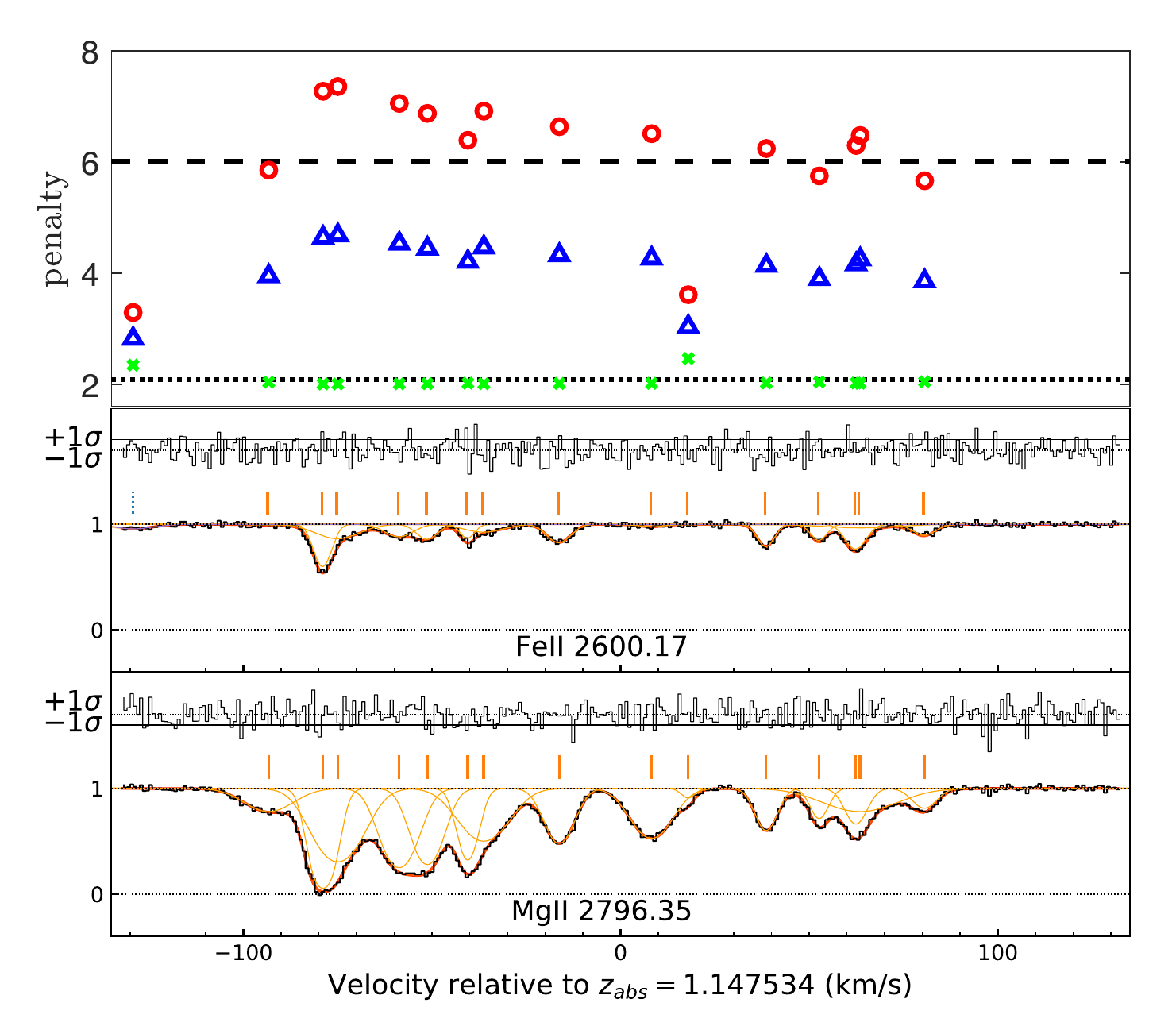}
\caption{Two example transitions from Region A. These are synthetic fiducial spectra, derived using SpIC$_{\textrm{H}}$. Upper panel: the penalty term in Eq.\eqref{eq:SpIC} for each velocity component. The AICc (black dotted line) and BIC (black dashed line) are constant irrespective of absorption line strength. SpIC$_{\textrm{A}}$ (green cross), SpIC$_{\textrm{H}}$ (blue triangle) and SpIC$_{\textrm{B}}$ (red circle) take into account line strengths. Lower two panels: FeII2600 and MgII2796 transitions. The signal to noise per pixel is 50. The red continuous line shows the fiducial (i.e. true) model. Normalised residuals are shown above each spectrum.
\label{fig:pen-line-regA}}
\end{figure*}

\begin{figure*}
\centering
\includegraphics[width=0.98\linewidth]{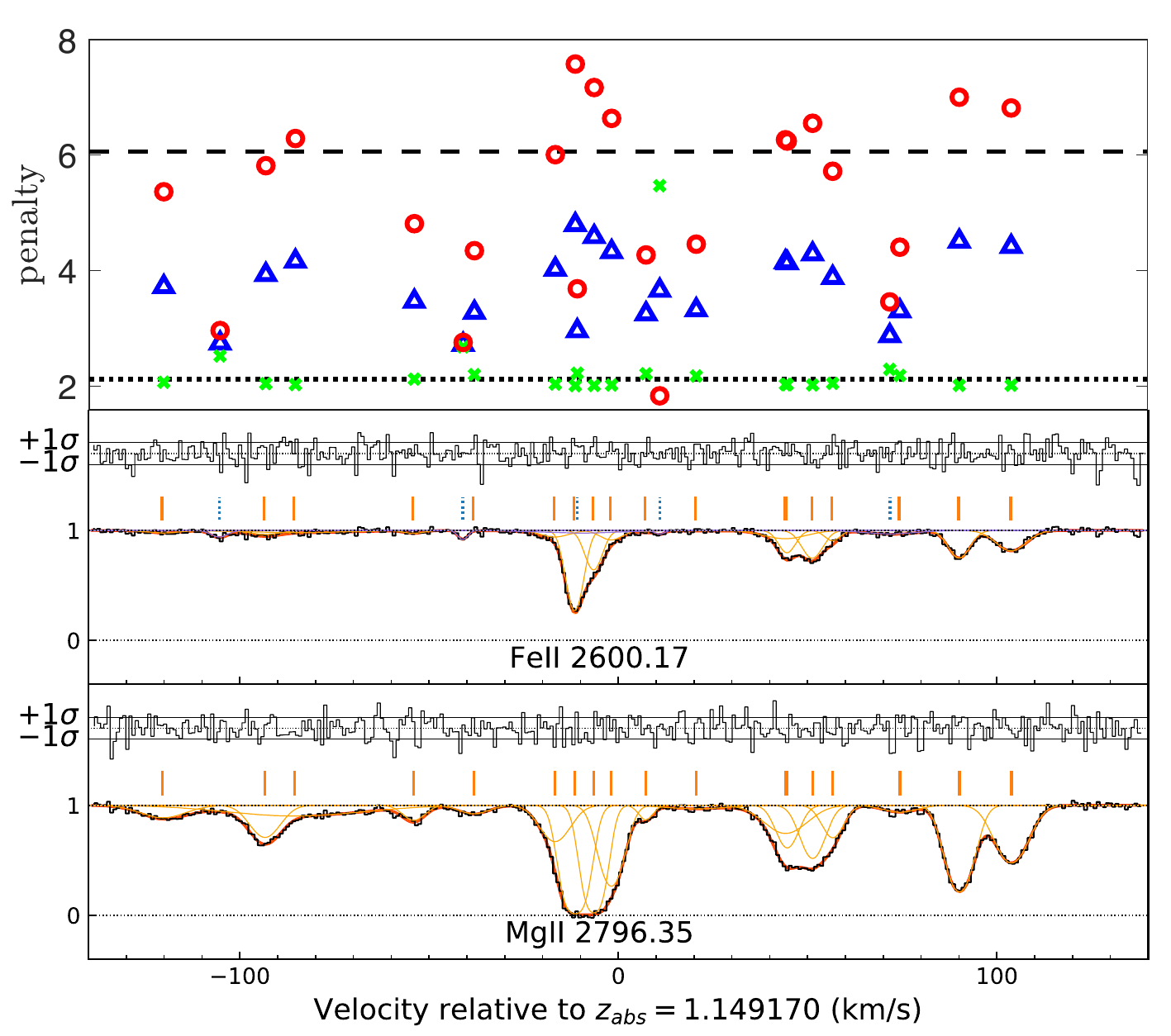}
\caption{Same as Fig.\ref{fig:pen-line-regA}, with BIC fiducial, for Region B.
\label{fig:pen-line-regB}}
\end{figure*}

\begin{figure*}
\centering
\includegraphics[width=0.98\linewidth]{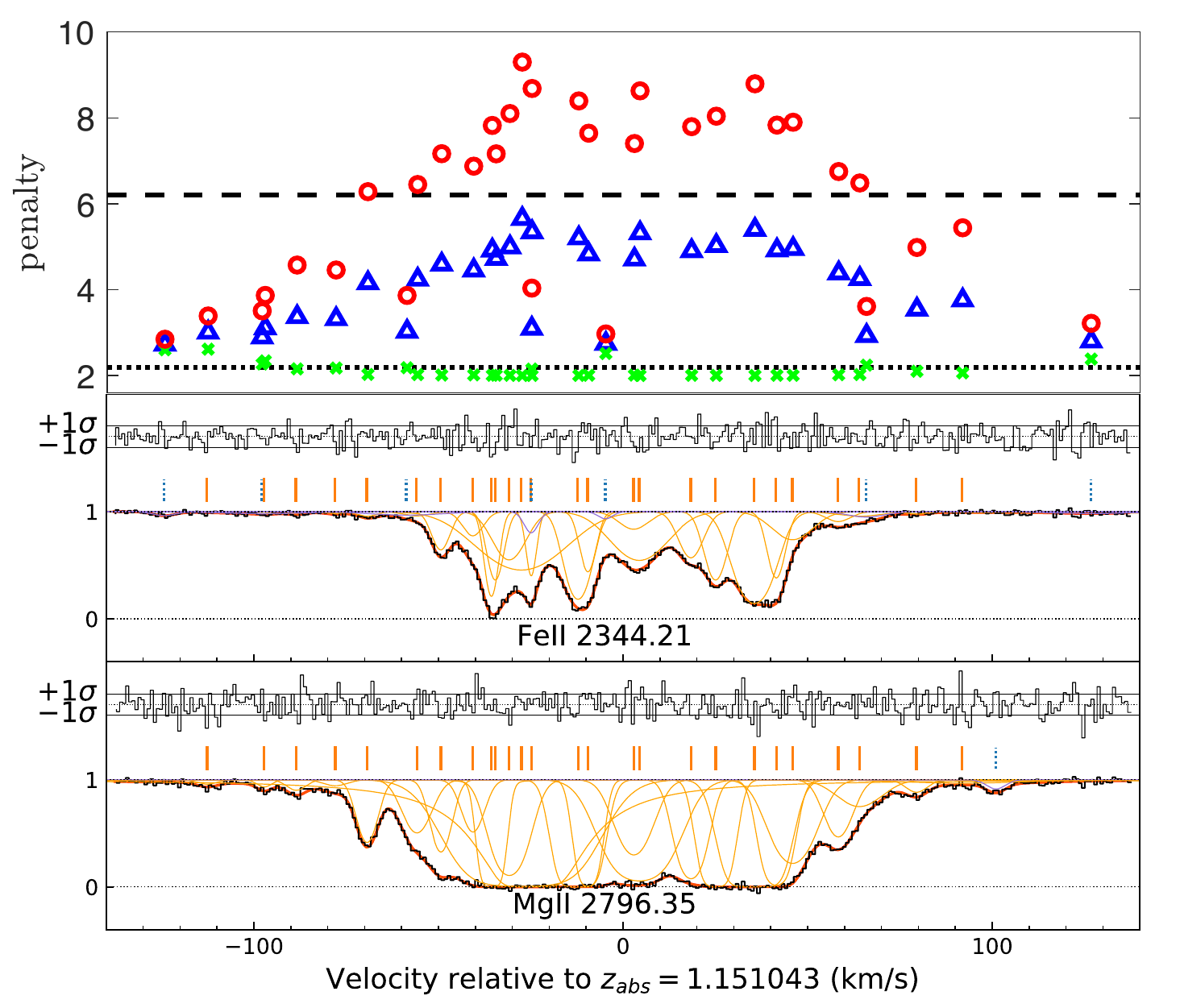}
\caption{Same as Fig.\ref{fig:pen-line-regA}, with AICc fiducial, for Region C.
\label{fig:pen-line-regC}}
\end{figure*}

Figures \ref{fig:pen-line-regA}, \ref{fig:pen-line-regB}, and \ref{fig:pen-line-regC} illustrate example transitions. Figure \ref{fig:pen-line-regA} shows FeII2600 and MgII2796 with the SpIC$_{\textrm{H}}$ fiducial model, Figure \ref{fig:pen-line-regB} shows FeII2600 and MgII2796 with the BIC fiducial model, and Figure \ref{fig:pen-line-regC} shows FeII2344 and MgII2796 with the AICc fiducial model. The S/N is 50. The fiducial models are shown as a continuous red line and individual velocity components illustrated with the thinner orange lines. Tick marks indicate the velocity component positions. The penalties for all 5 ICs are shown, AICc and BIC as horizontal lines as they are independent of line strength.

All three figures demonstrate that the SpIC$_{\textrm{H}}$ penalties generally fall between the AICc and BIC values and also between the SpIC$_{\textrm{A}}$ and SpIC$_{\textrm{B}}$ values, as intended. The SpIC$_{\textrm{B}}$ points exhibit the greatest scatter and the SpIC$_{\textrm{A}}$ points provide the smallest overall penalties. All three SpIC penalties show the dependence on line strength, most notable in Region C. The penalties for weak interlopers from SpIC$_{\textrm{B}}$ (e.g. +10 km/s in Fig. \ref{fig:pen-line-regB}) can be very small, suggesting that this IC is likely to generate spurious weak interlopers in the final model. SpIC$_{\textrm{A}}$ gives the same interloper a strong penalty, favouring strong lines.

To assess the relative performances of all 5 ICs, we fit all 6 sets of synthetic spectra using AI-VPFIT \citet{Lee2020AI-VPFIT}. Each dataset (incorporating transitions of MgI, MgII, FeII for Regions A and B, plus MnII for Region C) are fitted 5 times (for each of the two signal to noise ratios used), using AICc, BIC, SpIC$_{\textrm{A}}$, SpIC$_{\textrm{H}}$, and SpIC$_{\textrm{B}}$. The relevant IC is used at each generation to select the fittest model, to be passed on as the parent to the next generation. The choice of IC uniquely determines the way in which each generation evolves (see \citet{Lee2020AI-VPFIT}). Each of the final models are thus independently derived.  In addition, for comparative purposes, we also fit the fiducial spectrum, but using VPFIT \citep{VPFIT, web:VPFIT}, where the starting guesses are the true parameters. This is useful because it provides parameter estimates and uncertainties for each fiducial spectrum.

The database resulting from the analysis of the above numerical simulations allows us to evaluate the relative performances of each IC. The motivation behind the work described in this paper was to improve methods for measuring $\daa$. Therefore, we focus on this parameter of interest when comparing the relative performances of each IC, using the following four quantities for models and fiducials: bias (difference between the true and measured $\daa$), the variance on $\daa$, the number of fitted metal components, and the number of fitted interlopers. Tables \ref{tab:comparison_a}, \ref{tab:comparison_b}, and \ref{tab:comparison_c} present all the results, which are interpreted in Section \ref{Results}.

\section{Results} \label{Results}

We quantify the bias as follows:
\begin{equation} \label{eq:bsq_over_v}
\langle d_b^2 \rangle = \frac{1}{9} \sum_\mathrm{Region} \sum_\mathrm{Fiducial} \daa\rvert_{\mathrm{fid - obs}}^2 \, \,,
\end{equation}
where the first summation sign means a summation over the 3 regions (A, B, and C) and the second summation sign means a summation over the 3 fiducials within each region at the same S/N (hence the factor of 1/9 to form the mean value). 

The corresponding variance on $\daa$ is 
\begin{equation} \label{eq:variance}
\langle \sigma_{\!\alpha}^2 \rangle = \frac{1}{9} \sum_\mathrm{Region} \sum_\mathrm{Fiducial} \sigma_\alpha^2 \rvert_{\mathrm{obs}} \, \,,
\end{equation}
where $\sigma_\alpha$ is the uncertainty for the corresponding IC (e.g. for the AICc and S/N=50, we average the nine AICc $\sigma_{\alpha}$ entries in the last column in Tables \ref{tab:comparison_a}, \ref{tab:comparison_b} and \ref{tab:comparison_c}). We also compare the normalised $\chi^2_n = \chi^2/(N-k)$ values and the averaged number of free parameters associated with metal lines and interlopers, $\langle k_{\mathrm{metal}} \rangle$ and $\langle k_{\mathrm{int}} \rangle$. Table \ref{tab:summary} gives the quantities above, for each S/N, for each IC.

\subsection{S/N=50}

The upper part of Table \ref{tab:summary} shows that the preferred ICs at this S/N are AICc and SpIC$_{\textrm{H}}$. Given the uncertainties it is not possible to confidently say which of the two works better. SpIC$_{\textrm{A}}$ also performs reasonably well. BIC and SpIC$_{\textrm{B}}$ perform the worst. These conclusions can be reached as follows.

At S/N=50, the particular set of 9 fiducials happens to give $\langle \chi_n^2 \rangle = 1.000$. We thus expect the ICs all to have $\langle \chi_n^2 \rangle$ consistent with 1.000. Table \ref{tab:summary} shows that AICc, SpIC$_{\textrm{A}}$, SpIC$_{\textrm{H}}$ and SpIC$_{\textrm{B}}$ are consistent with 1, implying that these four ICs neither over-fit nor under-fit the data. However, $\langle \chi_n^2 \rangle$(BIC) is about 2.6$\sigma$ above unity. 

If we wish to follow the principle of parsimony, the preferred IC is that which simultaneously minimises $\langle d_b^2 \rangle$ and $\langle \sigma_{\alpha}^2 \rangle$. However, since these 2 quantities do not have the same means, nor the same mean uncertainties, we do not combine them into a single parameter and instead consider them separately.

The bias on $\daa$ obtained using AICc has the smallest value. SpIC$_{\textrm{A}}$ and SpIC$_{\textrm{H}}$ also yield a small bias and are consistent with each other within the empirical scatter. The values of $\langle d_b^2 \rangle$ from BIC and SpIC$_{\textrm{B}}$ models are more than 1$\sigma$ above the smallest AICc value. Noting that the measurement uncertainties are large, when considering for $\langle d_b^2 \rangle$ alone, AICc and SpIC$_{\textrm{H}}$ are tentatively favoured at S/N=50.

Now we consider the implications from $\langle \sigma_{\alpha}^2 \rangle$. We note that the measurement uncertainties are large so conclusions on the basis of $\langle \sigma_{\alpha}^2 \rangle$ alone are tentative. The S/N=50 results show that the smallest variances are given by BIC and SpIC$_{\textrm{B}}$. However, inspecting the corresponding values for $\langle d_b^2 \rangle$ and also $\langle k_{\mathrm{metal}} \rangle$, we see that $\langle \sigma_{\alpha}^2 \rangle$(BIC) and $\langle \sigma_{\alpha}^2 \rangle$(SpIC$_{\textrm{B}}$) are artificially small; $\daa$ is sensitive only to metal lines and fitting fewer metal line parameters naturally leads to smaller $\langle \sigma_{\alpha}^2 \rangle$, but this need not reflect reality (and indeed does not in this case, since we know the ``true'' value of $k_{\mathrm{metal}}$). Even if we ignore this fact, and base an interpretation {\it purely} on parsimony (i.e. we essentially ignore $\langle k_{\mathrm{metal}} \rangle$ and consider {\it only} $\langle d_b^2 \rangle$ and $\langle \sigma_{\alpha}^2 \rangle$), we still note the unduly large biases which would thus reject BIC and SpIC$_{\textrm{B}}$.

Finally we consider $\langle k_{\mathrm{metal}} \rangle$ and $\langle k_{\mathrm{int}} \rangle$. For $\langle k_{\mathrm{metal}} \rangle$ at S/N=50, AICc, SpIC$_{\textrm{A}}$, and SpIC$_{\textrm{H}}$ provide the closest match to the input fiducial. BIC and SpIC$_{\textrm{B}}$ underfit $\langle k_{\mathrm{metal}} \rangle$ relative to the input fiducial. BIC dramatically underfits $\langle k_{\mathrm{int}} \rangle$ and SpIC$_{\textrm{B}}$ dramatically overfits (because the latter penalises strong lines heavily but weak lines mildly). Whilst either underfitting or overfitting $\langle k_{\mathrm{int}} \rangle$ probably impacts very little on $\daa$, it is nevertheless undesirable.

\subsection{S/N=100}

The results from the synthetic spectra with S/N=100 are shown in the lower part of Table \ref{tab:summary}. In the following discussion we argue that SpIC$_{\textrm{H}}$ is the most effective IC at this S/N and that BIC also performs well.

We expect all ICs to have a $\langle \chi_n^2 \rangle$ consistent with 0.969 obtained from the 9 S/N=100 fiducial spectra. Whilst BIC has the highest $\langle \chi_n^2 \rangle$, all ICs are essentially consistent.

First consider only parsimony, i.e. let us again momentarily ignore $\langle k_{\mathrm{metal}} \rangle$ (and $\langle k_{\mathrm{int}} \rangle$) and look {\it only} at $\langle d_b^2 \rangle$ and $\langle \sigma_{\alpha}^2 \rangle$. Given the uncertainties, Table \ref{tab:summary} cannot distinguish between any of the ICs (even though BIC and SpIC$_{\textrm{B}}$ have the highest values). However, inspecting $\langle k_{\mathrm{metal}} \rangle$ and $\langle k_{\mathrm{int}} \rangle$ is telling. All of SpIC$_{\textrm{B}}$, AICc, and SpIC$_{\textrm{A}}$, dramatically overfit the number of interlopers. Whilst this effect is less obvious at S/N=50 in Table \ref{tab:summary} (the fiducial and AICc $\langle k_{\mathrm{int}} \rangle$ values seem to agree well), in fact AICc performs erratically in fitting interlopers (see e.g. Table \ref{tab:comparison_a}). On this basis one would therefore select SpIC$_{\textrm{H}}$ and BIC as the favoured ICs. 

\subsection{Which IC performs best overall?}

It is interesting that the S/N=50 and S/N=100 results reach slightly different conclusions. At S/N=50, the different ICs produce quite different solutions: overall, AICc and SpIC$_{\textrm{H}}$ are favoured whilst BIC and SpIC$_{\textrm{B}}$ are firmly rejected. On the other hand, at S/N=100, SpIC$_{\textrm{H}}$ and BIC are favoured whilst, if we regard it necessary to avoid unnecessary interlopers (which nevertheless do not appear to impact significantly on $\daa$ bias and variance, within the errors at least), then AICc, SpIC$_{\textrm{A}}$, and  SpIC$_{\textrm{B}}$ are disfavoured. If we do not care about including spurious parameters, we would conclude that all ICs perform well (at S/N=100 only). However, on balance, we suggest that SpIC$_{\textrm{H}}$ is the best ``all-rounder'' and is generally the most suited IC of those investigated, for the spectroscopic application considered.

\section{Conclusions}

We have carried out extensive spectral simulations of the $z_{abs}=1.15$ absorption system towards the quasar HE0515-0414. A new spectral information criterion, SpIC, has been developed and tested using VPFIT \citep{VPFIT, web:VPFIT} and AI-VPFIT algorithms \citep{Lee2020AI-VPFIT}. The purpose of SpIC is to (i) allow for highly localised parameter effects in the data being fitted and (ii) appropriately penalise according to absorption line strength.

By creating fiducial synthetic spectra using AICc, SpIC$_{\textrm{H}}$, and BIC, we generated three levels of model complexity (AICc having the most parameters, BIC the least). The synthetic fiducials (with independent noise characteristics and whose generating parameters are known) were then treated as real data and modelled using AI-VPFIT using the different ICs. The spectral simulation results allow us to quantify the relative performances of SpIC, AICc, and BIC.

Whilst the analysis described in this paper is somewhat complex, the conclusions are simple; at higher S/N, SpIC$_{\textrm{H}}$ performs better than the other ICs, including AICc, and is clearly preferred. At lower S/N, AICc most closely reproduces the input model, but if parsimony is followed, since AICc, SpIC$_{\textrm{A}}$ and SpIC$_{\textrm{H}}$ all provide good bias to variance ratios and good normalised chi-squared values with fewer parameters, these may be preferred. BIC does not suit this application and should be avoided.

Finally we note that, in modelling quasar absorption systems, it is impossible to recover the ``true'' underlying physical parameters. Although higher spectral resolution and S/N improve models, the intrinsic absorption line width ($b$-parameter) imposes a fundamental limitation. The true kinematic structure in quasar absorption systems is likely to be extremely complex. The purpose of model optimisation is thus merely to simultaneously minimise parameter variance and model bias and to quantify any remaining bias so that error budgets are properly understood. Identifying the most suitable information criteria, as we have done in this paper, is a necessary step towards that goal.

\section*{Acknowledgements}
CCL thanks the Royal Society for a Newton International Fellowship during the early stages of this work. JKW thanks the John Templeton Foundation, the Department of Applied Mathematics and Theoretical Physics and the Institute of Astronomy at Cambridge University for hospitality and support, and Clare Hall for a Visiting Fellowship during this work. We are grateful for supercomputer time using OzSTAR at the Centre for Astrophysics and Supercomputing at Swinburne University of Technology. Based on observations collected at the European Southern Observatory under ESO programme 102.A-0697(A).

\section*{Data Availability}
The analysis described in this paper is based on numerical simulations. All associated data files can be made available by the authors on request.

\bibliographystyle{mnras}
\bibliography{SpIC}

\begin{thebibliography}{}
\makeatletter
\relax
\def\mn@urlcharsother{\let\do\@makeother \do\$\do\&\do\#\do\^\do\_\do\%\do\~}
\def\mn@doi{\begingroup\mn@urlcharsother \@ifnextchar [ {\mn@doi@}
  {\mn@doi@[]}}
\def\mn@doi@[#1]#2{\def\@tempa{#1}\ifx\@tempa\@empty \href
  {http://dx.doi.org/#2} {doi:#2}\else \href {http://dx.doi.org/#2} {#1}\fi
  \endgroup}
\def\mn@eprint#1#2{\mn@eprint@#1:#2::\@nil}
\def\mn@eprint@arXiv#1{\href {http://arxiv.org/abs/#1} {{\tt arXiv:#1}}}
\def\mn@eprint@dblp#1{\href {http://dblp.uni-trier.de/rec/bibtex/#1.xml}
  {dblp:#1}}
\def\mn@eprint@#1:#2:#3:#4\@nil{\def\@tempa {#1}\def\@tempb {#2}\def\@tempc
  {#3}\ifx \@tempc \@empty \let \@tempc \@tempb \let \@tempb \@tempa \fi \ifx
  \@tempb \@empty \def\@tempb {arXiv}\fi \@ifundefined
  {mn@eprint@\@tempb}{\@tempb:\@tempc}{\expandafter \expandafter \csname
  mn@eprint@\@tempb\endcsname \expandafter{\@tempc}}}

\bibitem[\protect\citeauthoryear{{Akaike}}{{Akaike}}{1974}]{Akaike1974}
{Akaike} H.,  1974, IEEE Transactions on Automatic Control, \href
  {http://adsabs.harvard.edu/abs/1974ITAC...19..716A} {19, 716}

\bibitem[\protect\citeauthoryear{{Bainbridge} \& {Webb}}{{Bainbridge} \&
  {Webb}}{2017a}]{Bainbridge2017}
{Bainbridge} M.~B.,  {Webb} J.~K.,  2017a, \mn@doi [Universe]
  {10.3390/universe3020034}, \href
  {https://ui.adsabs.harvard.edu/#abs/2017Univ....3...34B} {3, 34}

\bibitem[\protect\citeauthoryear{{Bainbridge} \& {Webb}}{{Bainbridge} \&
  {Webb}}{2017b}]{gvpfit17}
{Bainbridge} M.~B.,  {Webb} J.~K.,  2017b, \mn@doi [MNRAS]
  {10.1093/mnras/stx179}, \href
  {http://adsabs.harvard.edu/abs/2017MNRAS.468.1639B} {468, 1639}

\bibitem[\protect\citeauthoryear{{Bhat} \& {Kumar}}{{Bhat} \&
  {Kumar}}{2010}]{Bhat2010}
{Bhat} H.~S.,  {Kumar} N.,  2010, On the derivation of the Bayesian Information
  Criterion, \url {https://faculty.ucmerced.edu/hbhat/BICderivation.pdf}

\bibitem[\protect\citeauthoryear{{Bozdogan}}{{Bozdogan}}{1987}]{Bozdogan1987}
{Bozdogan} H.,  1987, \mn@doi [Psychometrika] {10.1007/BF02294361}, 52, 345

\bibitem[\protect\citeauthoryear{Burnham \& Anderson}{Burnham \&
  Anderson}{2002}]{Burnham2002}
Burnham K.,  Anderson D.,  2002, Model selection and multimodel inference: a
  practical information-theoretic approach.
Springer Verlag

\bibitem[\protect\citeauthoryear{{Carswell} \& {Webb}}{{Carswell} \&
  {Webb}}{2014}]{VPFIT}
{Carswell} R.~F.,  {Webb} J.~K.,  2014, {VPFIT: Voigt profile fitting program},
  Astrophysics Source Code Library (\mn@eprint {ascl} {1408.015})

\bibitem[\protect\citeauthoryear{{Carswell} \& {Webb}}{{Carswell} \&
  {Webb}}{2020}]{web:VPFIT}
{Carswell} R.~F.,  {Webb} J.~K.,  2020, Bob Carswell's homepage, \url
  {https://people.ast.cam.ac.uk/~rfc/}

\bibitem[\protect\citeauthoryear{Einstein}{Einstein}{1954}]{Einstein1954}
Einstein A.,  1954, The Problem of Space, Ether, and the Field in Physics
  (1934). English translation: Ideas and Opinions.
Crown Publishers, Inc., New York

\bibitem[\protect\citeauthoryear{{Hurvich} \& {Tsai}}{{Hurvich} \&
  {Tsai}}{1989}]{Hurvich1989}
{Hurvich} C.~M.,  {Tsai} C.-L.,  1989, \mn@doi [Biometrika]
  {10.1093/biomet/76.2.297}, 76, 297

\bibitem[\protect\citeauthoryear{{Kotu{\v{s}}}, {Murphy}  \&
  {Carswell}}{{Kotu{\v{s}}} et~al.}{2017}]{Kotus2017}
{Kotu{\v{s}}} S.~M.,  {Murphy} M.~T.,   {Carswell} R.~F.,  2017, \mn@doi
  [MNRAS] {10.1093/mnras/stw2543}, \href
  {https://ui.adsabs.harvard.edu/abs/2017MNRAS.464.3679K} {464, 3679}

\bibitem[\protect\citeauthoryear{Laird}{Laird}{1919}]{Laird1919}
Laird J.,  1919, The Monist, 29, 321

\bibitem[\protect\citeauthoryear{{Lee}, {Webb}, {Carswell}  \&
  {Milakovic}}{{Lee} et~al.}{2020}]{Lee2020AI-VPFIT}
{Lee} C.-C.,  {Webb} J.~K.,  {Carswell} R.~F.,   {Milakovic} D.,  2020, arXiv
  e-prints, \href {https://ui.adsabs.harvard.edu/abs/2020arXiv200802583L} {p.
  arXiv:2008.02583}

\bibitem[\protect\citeauthoryear{{Liddle}}{{Liddle}}{2007}]{Liddle2007}
{Liddle} A.~R.,  2007, \mn@doi [\mnras] {10.1111/j.1745-3933.2007.00306.x},
  \href {https://ui.adsabs.harvard.edu/abs/2007MNRAS.377L..74L} {377, L74}

\bibitem[\protect\citeauthoryear{{Milakovi{\'c}}, {Lee}, {Carswell}, {Webb},
  {Molaro}  \& {Pasquini}}{{Milakovi{\'c}} et~al.}{2020}]{MilakovicHE05152020}
{Milakovi{\'c}} D.,  {Lee} C.-C.,  {Carswell} R.~F.,  {Webb} J.~K.,  {Molaro}
  P.,   {Pasquini} L.,  2020, arXiv e-prints, \href
  {https://ui.adsabs.harvard.edu/abs/2020arXiv200810619M} {p. arXiv:2008.10619}

\bibitem[\protect\citeauthoryear{Newton}{Newton}{1726}]{Newton1726}
Newton I.,  1726, in the {\it Principia}, English translation by Andrew Motte,
  Book III, Rules I and II, page 384.
Daniel Adee, New York, 1846

\bibitem[\protect\citeauthoryear{Portet}{Portet}{2020}]{Portet2020}
Portet S.,  2020, \mn@doi [Infectious Disease Modelling]
  {https://doi.org/10.1016/j.idm.2019.12.010}, 5, 111

\makeatother
\end{thebibliography}

\begin{table*}
\begin{tabular}{ c l c c c c c c c} 
\hline
Region &
Criterion &
S/N &
$\chi^2_{\mathrm{measured}}$ & 
$n_{\mathrm{metal}}$ & 
$n_{\mathrm{int}}$ & 
$k$ & 
$\left(\daa\right)$ &  
$\sigma_{\!\alpha}$ \\[0.5ex]  \hline
Region A & {\it fiducial} & {\it 50} & {\it 2358.75} & {\it 20} & {\it 6} & {\it 138} & {\it 5.84} & {\it 3.57} \\
 (AICc) & AICc          & 50  & 2334.86 & 18 & 9 & 135 & 5.62 & 3.34 \\
  & BIC                 & 50  & 2502.25 & 14 & 3 & 93  & 4.30 & 3.25 \\
  & SpIC$_{\textrm{A}}$ & 50  & 2384.69 & 17 & 6 & 120 & 6.55 & 3.54 \\
  & SpIC$_{\textrm{H}}$ & 50  & 2374.51 & 15 & 4 & 102 & 6.65 & 3.49 \\
  & SpIC$_{\textrm{B}}$ & 50  & 2400.84 & 16 & 9 & 123 & 4.76 & 3.17 \\ \hline
Region A & {\it fiducial} & {\it 100} & {\it 2362.38} & {\it 20} & {\it 6} & {\it 138} & {\it 5.60} & {\it 1.77} \\
 (AICc) & AICc          & 100 & 2325.67 & 19 & 11  & 147 & 5.85 & 1.78 \\
  & BIC                 & 100 & 2436.59 & 18 & 5  & 123 & 3.64 & 1.78 \\
  & SpIC$_{\textrm{A}}$ & 100 & 2355.13 & 19 & 8  & 138 & 6.29 & 1.78 \\
  & SpIC$_{\textrm{H}}$ & 100 & 2364.46 & 19 & 8  & 138 & 5.18 & 1.75 \\
  & SpIC$_{\textrm{B}}$ & 100 & 2372.26 & 18 & 12 & 144 & 5.53 & 1.79 \\ \hline \hline
Region A & {\it fiducial} & {\it 50} & {\it 2394.69} & {\it 15} & {\it 1} & {\it 93} & {\it 6.79} & {\it 3.00} \\
 (BIC) & AICc           & 50  & 2311.51 & 15 & 10 & 120 & 4.53 & 3.00 \\
  & BIC                 & 50  & 2434.69 & 14 & 1  & 87  & 8.94 & 3.07 \\
  & SpIC$_{\textrm{A}}$ & 50  & 2393.05 & 15 & 1  & 93  & 6.63 & 2.96 \\
  & SpIC$_{\textrm{H}}$ & 50  & 2374.51 & 15 & 3  & 99  & 8.87 & 3.07 \\
  & SpIC$_{\textrm{B}}$ & 50  & 2390.84 & 15 & 4  & 102 & 8.38 & 2.92 \\ \hline
Region A & {\it fiducial} & {\it 100} & {\it 2470.44} & {\it 15} & {\it 1} & {\it 93} & {\it 7.40} & {\it 1.52} \\
 (BIC) & AICc           & 100 & 2438.74 & 15 & 4  & 102 & 7.57 & 1.51 \\
  & BIC                 & 100 & 2471.28 & 15 & 1  & 93  & 7.78 & 1.53 \\
  & SpIC$_{\textrm{A}}$ & 100 & 2423.30 & 17 & 4  & 114 & 8.13 & 1.56 \\
  & SpIC$_{\textrm{H}}$ & 100 & 2462.87 & 15 & 1  & 93  & 7.59 & 1.53 \\
  & SpIC$_{\textrm{B}}$ & 100 & 2409.31 & 15 & 6  & 108 & 7.51 & 1.52 \\ \hline \hline
Region A & {\it fiducial} & {\it 50} & {\it 2439.00} & {\it 17} & {\it 6} & {\it 120} & {\it 4.13} & {\it 3.37} \\
 (SpIC$_{\textrm{H}}$) & AICc & 50 & 2454.17 & 16 & 6  & 114 & 2.52 & 3.44 \\
  & BIC                 & 50  & 2576.83 & 15 & 1  & 93  & 0.58 & 3.32 \\
  & SpIC$_{\textrm{A}}$ & 50  & 2453.87 & 17 & 5  & 117 & 1.03 & 3.58 \\
  & SpIC$_{\textrm{H}}$ & 50  & 2472.56 & 15 & 6  & 108 & 1.64 & 3.19 \\
  & SpIC$_{\textrm{B}}$ & 50  & 2432.86 & 16 & 8  & 120 & 2.41 & 3.18 \\ \hline
Region A & {\it fiducial} & {\it 100} & {\it 2267.61} & {\it 17} & {\it 6} & {\it 120} & {\it 4.03} & {\it 1.66} \\
 (SpIC$_{\textrm{H}}$) & AICc & 100 & 2261.76 & 17 & 7 & 123 & 3.71 & 1.66 \\
  & BIC                 & 100 & 2273.73 & 17 & 6  & 120 & 4.13 & 1.66 \\
  & SpIC$_{\textrm{A}}$ & 100 & 2291.19 & 17 & 6  & 120 & 3.27 & 1.70 \\
  & SpIC$_{\textrm{H}}$ & 100 & 2266.41 & 17 & 6  & 120 & 3.38 & 1.58 \\
  & SpIC$_{\textrm{B}}$ & 100 & 2240.90 & 17 & 9  & 129 & 3.95 & 1.64 \\ \hline \hline
\end{tabular}
\caption{Simulation results for Region A. The fiducial (shown in italics on the first line of each block as a reminder that parameters were derived using VPFIT, not AI-VPFIT) has been generated in 3 ways, using AICc, BIC, and SpIC$_\textrm{H}$ and at two S/N, hence there are 6 in total. Underneath each fiducial line, Column 1 identifies the region and in brackets shows the IC used to generate the fiducial. Column 2 shows the IC used to fit the fiducial. Column 3 gives the spectral signal to noise per pixel for the fiducial. Column 4 gives the overall $\chi^2$ for the entire fit (i.e. to all transitions/species). Columns 5 and 6 show the number of metal components and interlopers fitted. Column 7 gives the total number of free parameters in the fit. Columns 8 and 9 give the AI-VPFIT fitted $\daa$ and its estimated uncertainty, in units of $10^{-6}$. The results are discussed in detail in Section \ref{Results}.
\label{tab:comparison_a}}
\end{table*}

\begin{table*}
\begin{tabular}{ c l c c c c c c c} 
\hline
Region &
Criterion &
S/N &
$\chi^2_{\mathrm{measured}}$ & 
$n_{\mathrm{metal}}$ & 
$n_{\mathrm{int}}$ & 
$k$ & 
$\left(\daa\right)$ &  
$\sigma_{\!\alpha}$ \\[0.5ex]  \hline
Region B & {\it fiducial} & {\it 50} & {\it 2532.19} & {\it 18} & {\it 20} & {\it 168} & {\it 4.60} & {\it 4.58} \\
 (AICc) & AICc          & 50  & 2533.26 & 18 & 15 & 153 & 5.99 & 4.51 \\
  & BIC                 & 50  & 2809.04 & 15 & 1  & 93  & 5.35 & 4.37 \\
  & SpIC$_{\textrm{A}}$ & 50  & 2639.31 & 18 & 6  & 126 & 6.32 & 4.47 \\
  & SpIC$_{\textrm{H}}$ & 50  & 2620.48 & 18 & 8  & 132 & 5.19 & 4.49 \\
  & SpIC$_{\textrm{B}}$ & 50  & 2585.23 & 17 & 12 & 138 & 6.47 & 4.51 \\ \hline
Region B & {\it fiducial} & {\it 100} & {\it 2365.40} & {\it 18} & {\it 20} & {\it 168} & {\it 6.38} & {\it 2.21} \\
 (AICc) & AICc          & 100 & 2332.53 & 18 & 23 & 177 & 5.55 & 2.23 \\
  & BIC                 & 100 & 2401.13 & 18 & 18 & 162 & 6.13 & 2.24 \\
  & SpIC$_{\textrm{A}}$ & 100 & 2402.55 & 18 & 18 & 162 & 6.02 & 2.25 \\
  & SpIC$_{\textrm{H}}$ & 100 & 2360.10 & 18 & 23 & 177 & 5.56 & 2.24 \\
  & SpIC$_{\textrm{B}}$ & 100 & 2356.60 & 18 & 23 & 177 & 5.61 & 2.23 \\ \hline \hline
Region B & {\it fiducial} & {\it 50} & {\it 2516.23} & {\it 16} & {\it 1} & {\it 99} & {\it 6.17} & {\it 4.31} \\
 (BIC) & AICc           & 50  & 2487.11 & 16 & 5  & 111 & 5.56 & 4.26 \\
  & BIC                 & 50  & 2521.59 & 16 & 0  & 96  & 4.59 & 4.29 \\
  & SpIC$_{\textrm{A}}$ & 50  & 2515.80 & 16 & 1  & 99  & 4.71 & 4.30 \\
  & SpIC$_{\textrm{H}}$ & 50  & 2521.32 & 16 & 0  & 96  & 5.00 & 4.31 \\
  & SpIC$_{\textrm{B}}$ & 50  & 2538.17 & 14 & 11 & 117 & 2.25 & 4.38 \\ \hline
Region B & {\it fiducial} & {\it 100} & {\it 2520.31} & {\it 16} & {\it 1} & {\it 99} & {\it 6.03} & {\it 2.18} \\
 (BIC) & AICc           & 100 & 2462.75 & 16 & 6  & 114 & 5.26 & 2.15 \\
  & BIC                 & 100 & 2512.73 & 16 & 1  & 99  & 5.34 & 2.18 \\
  & SpIC$_{\textrm{A}}$ & 100 & 2488.92 & 18 & 10 & 138 & 5.28 & 2.17 \\
  & SpIC$_{\textrm{H}}$ & 100 & 2494.18 & 16 & 3  & 105 & 5.58 & 2.17 \\
  & SpIC$_{\textrm{B}}$ & 100 & 2452.15 & 16 & 9  & 123 & 5.05 & 2.16 \\ \hline \hline
Region B & {\it fiducial} & {\it 50} & {\it 2472.68} & {\it 18} & {\it 10} & {\it 138} & {\it 5.57} & {\it 4.24} \\
 (SpIC$_{\textrm{H}}$) & AICc & 50 & 2422.31 & 19 & 13 & 153 & 8.47 & 4.13 \\
  & BIC                 & 50  & 2609.87 & 17 & 1  & 105 & 9.72 & 4.28 \\
  & SpIC$_{\textrm{A}}$ & 50  & 2528.58 & 19 & 4  & 126 & 6.84 & 4.24 \\
  & SpIC$_{\textrm{H}}$ & 50  & 2451.01 & 19 & 9  & 141 & 6.08 & 4.15 \\
  & SpIC$_{\textrm{B}}$ & 50  & 2385.56 & 19 & 21 & 177 & 9.85 & 4.14 \\ \hline
Region B & {\it fiducial} & {\it 100} & {\it 2470.46} & {\it 18} & {\it 10} & {\it 138} & {\it 3.52} & {\it 2.21} \\
 (SpIC$_{\textrm{H}}$) & AICc & 100 & 2460.72 & 18 & 11 & 141 & 3.02 & 2.22 \\
  & BIC                 & 100 & 2490.36 & 18 & 9  & 135 & 2.95 & 2.30 \\
  & SpIC$_{\textrm{A}}$ & 100 & 2473.03 & 18 & 10 & 138 & 2.65 & 2.25 \\
  & SpIC$_{\textrm{H}}$ & 100 & 2474.06 & 18 & 10 & 138 & 2.80 & 2.29 \\
  & SpIC$_{\textrm{B}}$ & 100 & 2441.08 & 18 & 13 & 147 & 3.27 & 2.27 \\ \hline \hline
\end{tabular}
\caption{As Table \ref{tab:comparison_a} except for Region B.
\label{tab:comparison_b}}
\end{table*}

\begin{table*}
\begin{tabular}{ c l c c c c c c c} 
\hline
Region &
Criterion &
S/N &
$\chi^2_{\mathrm{measured}}$ & 
$n_{\mathrm{metal}}$ & 
$n_{\mathrm{int}}$ & 
$k$ & 
$\left(\daa\right)$ &  
$\sigma_{\!\alpha}$ \\[0.5ex]  \hline
Region C & {\it fiducial} & {\it 50} & {\it 2982.07} & {\it 26} & {\it 22} & {\it 240} & {\it 3.32} & {\it 4.20} \\
 (AICc) & AICc          & 50  & 3041.02 & 24 & 13 & 201 & 3.77 & 3.92 \\
  & BIC                 & 50  & 3361.46 & 20 & 1  & 139 & -1.59 & 2.75 \\
  & SpIC$_{\textrm{A}}$ & 50  & 3194.92 & 23 & 6  & 172 & -1.43 & 2.99 \\
  & SpIC$_{\textrm{H}}$ & 50  & 3177.60 & 22 & 5  & 163 & -1.81 & 2.86 \\
  & SpIC$_{\textrm{B}}$ & 50  & 3215.00 & 18 & 10 & 152 & 1.55 & 2.94 \\ \hline
Region C & {\it fiducial} & {\it 100} & {\it 2740.12} & {\it 26} & {\it 22} & {\it 240} & {\it 5.76} & {\it 1.98} \\
 (AICc) & AICc          & 100 & 2757.52 & 27 & 22 & 247 & 6.61 & 2.07 \\
  & BIC                 & 100 & 2867.49 & 26 & 16 & 222 & 5.37 & 2.05 \\
  & SpIC$_{\textrm{A}}$ & 100 & 2740.22 & 27 & 21 & 244 & 7.18 & 2.14 \\
  & SpIC$_{\textrm{H}}$ & 100 & 2802.28 & 27 & 20 & 241 & 5.14 & 2.13 \\
  & SpIC$_{\textrm{B}}$ & 100 & 2796.10 & 24 & 22 & 228 & 3.87 & 2.10 \\ \hline \hline
Region C & {\it fiducial} & {\it 50} & {\it 2997.81} & {\it 22} & {\it 3}  & {\it 157} & {\it 5.84} & {\it 3.30} \\
 (BIC) & AICc           & 50  & 2931.90 & 22 & 9  & 175 & 1.31 & 3.17 \\
  & BIC                 & 50  & 3098.66 & 20 & 1  & 139 & 2.49 & 3.14 \\
  & SpIC$_{\textrm{A}}$ & 50  & 2988.95 & 22 & 3  & 157 & 2.71 & 3.20 \\
  & SpIC$_{\textrm{H}}$ & 50  & 2977.54 & 22 & 4  & 160 & 3.76 & 3.28 \\
  & SpIC$_{\textrm{B}}$ & 50  & 2871.52 & 22 & 19 & 204 &-0.10 & 3.20 \\ \hline
Region C & {\it fiducial} & {\it 100} & {\it 2899.76} & {\it 22} & {\it 3}  & {\it 157} & {\it 6.22} & {\it 1.68} \\
 (BIC) & AICc           & 100 & 2872.59 & 22 & 6  & 166 & 4.90 & 1.64 \\
  & BIC                 & 100 & 2900.30 & 22 & 3  & 157 & 6.35 & 1.74 \\
  & SpIC$_{\textrm{A}}$ & 100 & 2832.88 & 22 & 11 & 181 & 6.16 & 1.70 \\
  & SpIC$_{\textrm{H}}$ & 100 & 2874.80 & 22 & 5  & 163 & 5.27 & 1.73 \\
  & SpIC$_{\textrm{B}}$ & 100 & 2847.37 & 22 & 8  & 172 & 8.05 & 1.73 \\ \hline \hline
Region C & {\it fiducial} & {\it 50}  & {\it 2954.00} & {\it 24} & {\it 11} & {\it 194} & {\it 3.28} & {\it 3.55} \\
 (SpIC$_{\textrm{H}}$) & AICc & 50 & 2947.41 & 24 & 8 & 185 & -0.23 & 3.21 \\
  & BIC                 & 50  & 3151.02 & 23 & 0  & 154 & -0.14 & 3.05 \\
  & SpIC$_{\textrm{A}}$ & 50  & 3024.23 & 24 & 3  & 170 & -1.26 & 3.27 \\
  & SpIC$_{\textrm{H}}$ & 50  & 3032.51 & 24 & 3  & 170 & -0.94 & 3.25 \\
  & SpIC$_{\textrm{B}}$ & 50  & 3007.39 & 23 & 11 & 186 & -0.08 & 2.86 \\ \hline 
Region C & {\it fiducial} & {\it 100} & {\it 2792.67} & {\it 24} & {\it 11} & {\it 194} & {\it 4.67} & {\it 1.71} \\
 (SpIC$_{\textrm{H}}$) & AICc & 100 & 2750.74 & 24 & 14 & 203 & 2.06 & 1.77 \\
  & BIC                 & 100 & 2874.84 & 24 & 7  & 182 & 2.01 & 1.72 \\
  & SpIC$_{\textrm{A}}$ & 100 & 2768.77 & 24 & 12 & 197 & 2.29 & 1.74 \\
  & SpIC$_{\textrm{H}}$ & 100 & 2811.04 & 24 & 10 & 191 & 1.95 & 1.76 \\
  & SpIC$_{\textrm{B}}$ & 100 & 2833.15 & 24 & 13 & 199 & 1.40 & 1.66 \\ \hline \hline
\end{tabular}
\caption{As Table \ref{tab:comparison_a} except for Region C.
\label{tab:comparison_c}}
\end{table*}

\begin{table*}
\begin{tabular}{ c c c c c c c} 
\hline 
 & \multicolumn{5}{c}{$S/N = 50$} \\ \cmidrule(lr){2-6}
\textbf{Criterion} & 
\textbf{$\langle \chi^2_n \rangle$} & 
\textbf{$\langle d_b^2 \rangle$} & 
\textbf{$\langle \sigma_{\!\alpha}^2 \rangle$} &
\textbf{$\langle k_{\mathrm{metal}} \rangle$} & 
\textbf{$\langle k_{\mathrm{int}} \rangle$} \\[0.5ex] 
 \hline
{\it fiducial} & {\it 1.000 $\pm$ 0.009} & - & {\it 14.64 $\pm$ 1.40} & {\it 123.0} & {\it 26.7} \\
AICc                & $0.992 \pm 0.011$ & $5.72 \pm 2.31$ & $13.70 \pm 1.37$ & 120.3 & 29.3 \\
BIC                 & $1.044 \pm 0.017$ & $9.66 \pm 2.62$ & $12.62 \pm 1.53$ & 108.0 &  3.0 \\
SpIC$_{\textrm{A}}$ & $1.013 \pm 0.013$ & $7.76 \pm 2.88$ & $13.38 \pm 1.44$ & 119.4 & 11.7 \\
SpIC$_{\textrm{H}}$ & $1.007 \pm 0.013$ & $6.85 \pm 3.04$ & $13.03 \pm 1.47$ & 116.1 & 14.0 \\
SpIC$_{\textrm{B}}$ & $1.006 \pm 0.013$ & $10.39 \pm 3.75$ & $12.49 \pm 1.64$ & 111.6 & 35.0 \\
\hline
 & \multicolumn{5}{c}{$S/N = 100$}  \\ \cmidrule(lr){2-6}
\textbf{Criterion} & 
\textbf{$\langle \chi^2_n \rangle$} & 
\textbf{$\langle d_b^2 \rangle$} & 
\textbf{$\langle \sigma_{\!\alpha}^2 \rangle$} &
\textbf{$\langle k_{\mathrm{metal}} \rangle$} & 
\textbf{$\langle k_{\mathrm{int}} \rangle$} \\[0.5ex] 
 \hline
{\it fiducial} & {\it 0.969 $\pm$ 0.007} & - & {\it 3.60 $\pm$ 0.34} & {\it 123.0} & {\it 26.7} \\
AICc                & $0.963 \pm 0.007$ & $1.22 \pm 0.72$ & $3.65 \pm 0.35$ & 123.1 & 34.7 \\
BIC                 & $0.981 \pm 0.008$ & $1.34 \pm 0.82$ & $3.72 \pm 0.37$ & 121.7 & 22.0 \\
SpIC$_{\textrm{A}}$ & $0.968 \pm 0.005$ & $1.19 \pm 0.59$ & $3.76 \pm 0.36$ & 125.8 & 33.3 \\
SpIC$_{\textrm{H}}$ & $0.971 \pm 0.007$ & $1.19 \pm 0.78$ & $3.72 \pm 0.38$ & 123.1 & 28.7 \\
SpIC$_{\textrm{B}}$ & $0.966 \pm 0.007$ & $2.14 \pm 1.17$ & $3.68 \pm 0.37$ & 120.2 & 38.3 \\
\hline
\end{tabular}
\caption{Summary of results. 
The fiducial is shown in italics on the first line of each block as a reminder that parameters were derived using VPFIT, not AI-VPFIT. Underneath the fiducial line, Column 1 indicates the IC used to fit the fiducial. Column 2 gives the averaged normalised $\chi^2$ value. Columns 3 and 4 are calculated using Eqs. \eqref{eq:bsq_over_v} and \eqref{eq:variance}. Columns 5 and 6 give the average number of parameters for metals and interlopers respectively. The results are discussed in Section \ref{Results}.
\label{tab:summary}}
\end{table*}

\bsp
\label{lastpage}
\end{document}